\documentclass[journal]{IEEEtran}
\IEEEoverridecommandlockouts
\usepackage{cite}
\usepackage{amsmath,amssymb,amsfonts}
\usepackage{algorithmic}
\usepackage{graphicx}
\usepackage{textcomp}
\usepackage{xcolor}
\usepackage{multirow}
\usepackage{enumitem}
\usepackage{bm}
\usepackage{comment}
\newcommand{\argmax}[1]{\underset{#1}{\operatorname{arg}\,\operatorname{max}}\;}

\def\BibTeX{{\rm B\kern-.05em{\sc i\kern-.025em b}\kern-.08em
    T\kern-.1667em\lower.7ex\hbox{E}\kern-.125emX}}
\begin{document}



\title{Reconfigurable Low-Complexity Architecture for High Resolution Doppler Velocity Estimation in Integrated Sensing and Communication System}

\author{
\IEEEauthorblockN{Aakanksha Tewari$^{\ast}$, Samarth Sharma Bhardwaj$^{\dagger}$, Sumit~J~Darak$^{\ast}$, Shobha~Sundar~Ram$^{\ast}$}
\IEEEauthorblockA{$^{\ast}$Indraprashtha Institute of Information Technology Delhi, New Delhi, India \\$^{\dagger}$Indian Institute of Technology Kanpur, Kanpur, Uttar Pradesh 208016 India\\
E-mail:\{aakankshat,sumit,shobha\}@iiitd.ac.in, samarthsb23@iitk.ac.in}
}




\maketitle
\begin{abstract}
In millimeter-wave integrated sensing and communication (ISAC) systems for intelligent transportation, radar and communication share spectrum and hardware in a time-division manner. Radar rapidly detects and localizes mobile users (MUs), after which communication proceeds through narrow beams identified by radar. Achieving fine Doppler resolution for MU–clutter discrimination requires long coherent processing intervals, reducing communication time and throughput. To address this, we propose a reconfigurable architecture for Doppler estimation realized on a system-on-chip using hardware–software co-design. The architecture supports algorithm-level reconfiguration, dynamically switching between low-complexity, high-speed FFT-based coarse estimation and high-complexity ESPRIT-based fine estimation. We introduce modifications to ESPRIT that achieve 6.7$\times$ faster execution while reducing memory and multiplier usage by 79\% and 63\%, respectively, compared to state-of-the-art approaches, without compromising accuracy. Additionally, the reconfigurable architecture can switch to lower slow-time packets under high-SNR conditions, improving latency further by 2$\times$ with no loss in performance.
\end{abstract}


\section{Introduction} 
Millimeter-wave (mmW) Integrated Sensing and Communication (ISAC) systems have emerged as key enablers for next-generation intelligent transportation systems (ITS), digital twin, and Industrial Internet of Things (IIoT) \cite{kvm2019mmwjrc,8999605,10870127, ISAC_intro3}. These applications require high-speed, accurate radar sensing to rapidly localize mobile users (MUs) for subsequent high–data-rate, low-latency communication. Co-designed ISAC systems deliver radar sensing and communication on a common spectrum, waveform, and hardware, promoting infrastructure reuse and cost effectiveness. A notable example is the mmW IEEE 802.11ad-based ISAC in \cite{duggal2020doppler, 8835525, sneh2022ieee, kumari2017ieee}. Here, the assumption is that the MU for communication is a mobile target first sensed by the radar. Thus, the primary role of radar signal processing (RSP) is to provide low-latency, accurate localization of targets within the field of view, enabling directional, high-gain communication and improved overall system performance.

\begin{figure}[h]
    \centering
     \includegraphics[scale = 0.19]{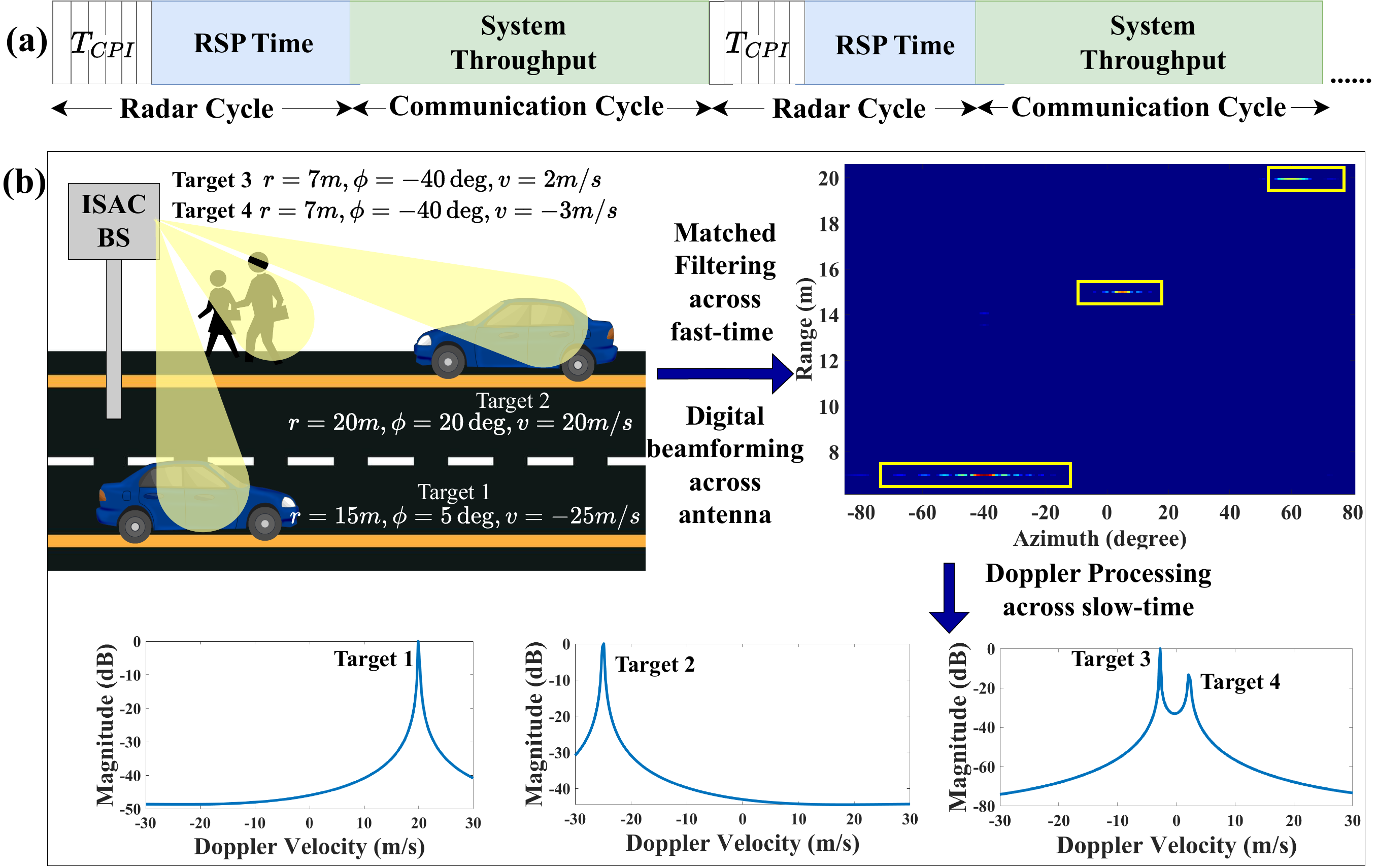}
    \caption{(a) ISAC with TDM between radar and communication functionalities (b) ISAC system with BS supporting 3D RSP and multiple MUs }
    \label{fig:sys_fig}
\end{figure}

In ISAC systems, radar sensing and communication is carried out either simultaneously or in a time-division multiplexing (TDM) manner. While simultaneous operation offers theoretical advantages, it poses significant challenges related to interference management, synchronization, and the requirement of full-duplex radios \cite{li2017joint}. Hence, the TDM approach is often considered more practical, where a radar cycle is followed by a communication cycle, as shown in Figure \ref{fig:sys_fig}(a). In this setting, the objective is to complete radar sensing and signal processing as quickly as possible, thereby leaving more time for communication and improving throughput. The radar cycle consists of two main steps: (1) coherent processing interval (CPI), which comprises the transmission of multiple pulses during which target-scattered returns are gathered, (2) radar signal processing (RSP), during which targets are detected and localized in range, azimuth, and Doppler. Ideally, both the sensing and processing steps should be of very short duration.

Existing works \cite{tewari2024reconfigrsp, sneh2022ieee} typically localize targets in the range–azimuth domain first, and then discriminate between tightly spaced MUs and static clutter using high-resolution Doppler velocity estimation algorithms. This is highlighted in Figure \ref{fig:sys_fig}(b), where only 3 of 4 targets are resolvable in the range-azimuth domain; however, a high-resolution Doppler estimation can distinguish tightly spaced targets. When the conventional fast Fourier transform (FFT) is adopted for Doppler estimation, the velocity resolution is poor, but the processing time is short due to its low complexity architecture, as discussed in \cite{FFTvsMUSIC_TAES, tewari2024reconfigrsp}. On the other hand, subspace algorithms offer super-resolution with lower CPI. However, due to the complexity of the algorithms, the processing time is high. This work proposes a reconfigurable Doppler velocity estimation architecture that addresses these limitations.

Subspace-based methods such as multiple signal classification (MUSIC), estimation of signal parameters via rotational invariance techniques (ESPRIT), and their extensions have been investigated in theory and on hardware due to their super-resolution capabilities \cite{lee2018weighted, FFTvsMUSIC_TAES, 5545202,mansi2022spatialsensing, tewari2024reconfigrsp, subspaceNet_2025,esprit_hardware_sensor_2025, 11181132}. Here, eigenvalue decomposition (EVD) is a crucial step that is computationally intensive and significantly impacts Doppler estimation accuracy. Works \cite{MUSIC_TCAS1} and \cite{zhang2023HW_EVD_tvt} present hardware-efficient MUSIC architectures using Jacobi’s method for singular value decomposition (SVD) for EVD. Though SVD offers good resolution, the hardware design becomes highly intensive for large inputs. \cite{mansi2022spatialsensing} implements MUSIC on FPGA utilizing the Xilinx inbuilt QR factorization (QRF) IP core for EVD, though its accuracy is limited and restricted to square matrices. In prior works, FPGA-based ESPRIT is more complex than MUSIC due to an additional EVD parameter estimation and pseudo-inverse computation. \cite{esprit_hardware_sensor_2025} explores an efficient fixed-point ESPRIT for reduced hardware complexity. While \cite{subspaceNet_2025} shows that augmenting subspace methods with deep learning can enhance performance, the FPGA complexity increases due to extra processing blocks. Moreover, none of these works support run-time reconfigurability for selecting the appropriate algorithm/CPI for optimum Doppler performance under varying channel conditions, enabling enhanced ISAC metrics.

This work presents a reconfigurable Doppler velocity estimation architecture and its efficient realization on the Zynq multi-processor system on chip (MPSoC) platform via hardware–software co-design. The proposed architecture supports algorithm-level reconfiguration, enabling seamless runtime switching between low-complexity, high-speed coarse estimation based on FFT and high-complexity, high-accuracy fine estimation using subspace-based ESPRIT algorithms. Furthermore, we develop a low-complexity ESPRIT architecture that achieves comparable performance while maintaining significantly lower complexity compared to conventional ESPRIT and MUSIC implementations. The dynamic control over Doppler resolution enables SNR-based switching to fewer slow-time packets, further providing twice faster latency while maintaining optimal performance in separating close targets.

The paper is organized as follows: Section \ref{Sec:Doppler_algo} describes the subspace Doppler processing algorithms; Section \ref{Sec:reconfig_arch} explains the reconfigurable Zynq MPSoC architecture for Doppler estimation; Section \ref{Sec:performance_analysis} presents performance and hardware complexity analysis. Section \ref{Sec:conclusion} concludes the paper.

\section{Doppler Estimation Algorithms for ISAC}
\label{Sec:Doppler_algo}
We consider a stationary mmW ISAC base station (BS) equipped with an RSP transceiver and $Z$ MUs/radar targets in the environment. The BS uses the IEEE 802.11ad protocol for ISAC, with RSP and communication functionalities conducted in a TDM manner as discussed in \cite{sneh2022ieee, tewari2024reconfigrsp}. During the radar cycle, $N$ radar pulses, each comprising a Doppler-resilient Golay sequence \cite{duggal2020doppler, kumari2017ieee}, are transmitted at intervals of $T_{PRI}$, omnidirectionally from the BS transmitter. Radar echoes reflected from multiple targets impinge on a $Q$-element uniform linear array (ULA) at the BS receiver. After downconversion and digitization, the 3D radar data cube $\mathbf{X} \in \mathbb{C}^{M \times Q \times N}$ is obtained across $M$ fast-time samples, $Q$ antenna elements, and $N$ slow-time samples. The targets are localized across range-azimuth first, followed by Doppler estimation on each detection to differentiate MUs from static clutter, as shown.
\subsection{2D Range-Azimuth Localization}
 Each of the $N$ 2D packet in $\mathbf{X}$ is processed along the $M$ dimension through matched filtering for range estimation, and the $Q$ dimension through digital beamforming across $I$ search angles for the angle-of-arrival estimation, to obtain the range-azimuth ambiguity as shown,
 \par\noindent\small
\begin{align}
\mathbf{Y}[r,\phi,n]=\sum_{{z=1}}^Z a_z \mathbf{\Omega}[r-r_z,\phi-\phi_z] e^{-j\frac{4\pi}{\lambda}v_{z} n T_{PRI}}\in\mathbb{C}^{M\times I \times N}
\label{eq:RA_ambiguity}
\end{align}
\normalsize
 Here, $r_{z}$, $\phi_{z}$, $v_{z}$ are the range, azimuth and Doppler velocity of the $z^{th}$ target, $T_{PRI}$ is the pulse repetition interval (PRI), $\lambda$ is the mmW wavelength and $\mathbf{\Omega}\in \mathbb{C}^{M\times I}$ is the 2D sinc function in range-azimuth domain. Peak search on $\mathbf{Y}$ provides the range-azimuth estimates of the strongest target in the environment, as $<\hat{r}_z,\hat{\phi}_z>=\argmax{\phi,r}|\mathbf{Y}|$. 
\subsection{Doppler Velocity Estimation}
Post range-azimuth localization, the slow time vector, $\mathbf{y}\in \mathbb{C}^{N\times 1}$, is created by selecting the samples corresponding to the peak index, $(\hat{r}_z,\hat{\phi}_z)$ from each of the $N$ packets in $\mathbf{Y}$, $\mathbf{y}[n]= \mathbf{Y}[\hat{r}_z,\hat{\phi}_z,n]$. The CPI for Doppler processing is $T_{CPI}=NT_{PRI}$. Considering a special case of having $K$ targets in the detected range-azimuth bin, $\mathbf{y}$ is processed for Doppler velocity estimation using the following methods.
\subsubsection{Subspace Methods}
Subspace Doppler estimation methods offer super-resolution and can distinguish between two separate targets with fewer slow-time samples compared to the FFT. The detailed steps for ESPRIT are explained below.
\noindent
\begin{enumerate}[wide, labelwidth=!,labelindent=0pt,label=\Alph*]
\item[a)]{Spatial smoothening (SS) and autocovariance generation (ACG):} The input $\mathbf{y}$ is split into multiple $L<N$ length vectors, $\mathbf{s}_l=\mathbf{y}[l:l+L-1] \in \mathbb{C}^{L\times 1}$, where index $l$ spans, $l\in[0,1,\cdots,L-1]$. The averaged covariance matrix, $\mathbf{A}\in \mathbb{C}^{L\times L}$ is generated to minimize coherency between multiple targets as shown,
\par\noindent\small
\begin{align}
\mathbf{A} = \sum_{{l=0}}^{N-L-1} \mathbf{s}_l{\mathbf{s}_l^H}
\label{eq:ACG}
\end{align}
\normalsize
\item[b)]{Eigen vector decomposition (EVD):} The averaged covariance undergoes QR factorization, $\mathbf{A} = \mathbf{Q}\mathbf{R}$ and is decomposed into an orthogonal matrix $\mathbf{Q} \in \mathbb{C}^{L \times L}$ containing the eigen vectors of $\mathbf{A}$, and an upper traingulat matrix $\mathbf{R} \in \mathbb{C}^{L \times L}$ with diagonal elements corresponding to the eigen values of $\mathbf{A}$.  Eigen vectors of $\mathbf{Q}$ corresponding to $K$ large eigen values form the signal subspace, $\mathbf{E}=\mathbf{Q}[0:L,0:K-1]\in \mathbb{C}^{L\times K}$. 
\item[c)]{Pseudo inverse and eigen value calculation:} $\mathbf{E}$ is split into two signal subspace submatrices as, $\mathbf{E_1}=\mathbf{E}[0:L-2,0:K-1]\in \mathbb{C}^{(L-1)\times K}$, $\mathbf{E_2}=\mathbf{E}[1:L-1,0:K-1]\in \mathbb{C}^{(L-1)\times K}$.
$\mathbf{E_1}$ and $\mathbf{E_2}$ satisfy $\mathbf{E_1} = \bm{\epsilon} \mathbf{E_2}$, where, each of the $K$ eigen values of matrix $\bm{\epsilon}\in\mathbb{C}^{K \times K}$ are of the form $e^{-j\frac{4\pi}{\lambda}{v_k}T_{PRI}}$.
$\bm{\epsilon}$ is computed by taking the Moore–Penrose inverse or pseudo-inverse of the rectangular matrix $\mathbf{E_2}$ as shown,
\vspace{-0.2cm}
\begin{align}
\bm{\epsilon}=\mathbf{E_1} {\mathbf{E_2}^+}
\label{eq:epsilon_esprit}
\end{align}
The pseudo-inverse, $\mathbf{E_2}^+ \in \mathbb{C}^{K \times (L-1)}$ can be derived using singular value decomposition (SVD) as shown,   
\begin{align}\small
\mathbf{E_2} = \mathbf{U}\mathbf{\Sigma}\mathbf{V}^H
\label{eq:svd}
\end{align}
\vspace{-0.9cm}
\begin{align}\small
{\mathbf{E_2}}^+ = \mathbf{V}\mathbf{\Sigma}^+\mathbf{U}^H
\label{eq:svd_inv}
\end{align}
Here, $\mathbf{U}\in\mathbb{C}^{(L-1)\times(L-1)}$ and $\mathbf{V}\in \mathbb{C}^{K\times K}$ are unitary matrices satisfying $\mathbf{U}^{+}=\mathbf{U}^H$. $\mathbf{\Sigma}\in \mathbb{C}^{(L-1)\times K}$ is a rectangular diagonal matrix so $\mathbf{\Sigma}^+ \in \mathbb{C}^{K \times (L-1)}$ can be easily computed by taking the reciprocal of each diagonal element of $\mathbf{\Sigma}^T$. The SVD method requires high computational complexity; therefore, for our specific ISAC case, with $K$ MUs in a given range-azimuth cell, expected to be very small, we adopt a lower-complexity pseudo-inverse calculation as shown.
\
\begin{align}\small
{\mathbf{E_2}}^+ = (\mathbf{E_2}^H\mathbf{E_2})^+\mathbf{E_2}^H
\label{eq:custom_inv}
\end{align}
This is followed by finding $\bm{\epsilon}$ as shown in equation \eqref{eq:epsilon_esprit}. The Doppler velocity of each of the $K$ targets can be found from $\mu_k$, the eigen values of $\bm{\epsilon}$, as $\hat{v}_k=\angle(\frac{-\mu_k \lambda}{4\pi T_{PRI}})$.
\end{enumerate}
\subsubsection{Fast Fourier transform}
A zero-padded $P$-point FFT of $\mathbf{y}$ produces
$\tilde{\mathbf{y}}=\text{FFT}P([\mathbf{y},\mathrm{o}^{P-N}])$, and the Doppler
velocity estimate is obtained via peak search:
$\langle \hat{v}k \rangle = \arg\max{v} |\tilde{\mathbf{y}}|$.
The Doppler velocity resolution, i.e., the minimum velocity separation between
two targets, is $\Delta v^{\mathrm{res}}=\frac{\lambda}{2T{\mathrm{CPI}}}$.
While FFT-based processing is computationally light, it requires large $N$ for good Doppler resolution \cite{tewari2024reconfigrsp}.
For fixed $N$, the processing precision
$\Delta v^{\mathrm{pre}}=\frac{\lambda}{2PT_{\mathrm{PRI}}}$ can be made finer by
increasing the transform size $P$, though it does not improve resolution.


\section{Proposed Reconfigurable Architecture}
This section presents the hardware mapping and integration of FFT and ESPRIT algorithms on Zynq MPSoC.
\subsection{Hardware Architecture of ESPRIT}
Figure \ref{fig:esprit_arch} details the hardware architecture of ESPRIT. 
The SS and ACG steps involve BRAM partitioning for parallelizing the multiple covariance calculations $\mathbf{s}_l\mathbf{s}_l^H$ for faster averaged covariance computation $\mathbf{A}$. EVD is realized using the built-in AMD Xilinx QRF library. This is followed by matrix slicing to generate the signal subspace, followed by sub-matrix split to generate $\mathbf{E_1}$ and $\mathbf{E_2}$, stored in BRAMs A and B, respectively. The next step is the pseudo-inverse computation. We present two architectures for the same-
\subsubsection{SVD-based pseudo inverse}
This is discussed in equation \eqref{eq:svd_inv} of Section \ref{Sec:Doppler_algo}, and presented in Figure \ref{fig:esprit_arch}(b). $\mathbf{E_2}\in\mathbb{C}^{(L-1)\times K}$ is read from BRAM B and sent to the built-in AMD Xilinx SVD library. Since the SVD library only works on square matrices, a $(L-1)\times (L-1)$ matrix is streamed as input to SVD after appending zeroes to $\mathbf{E_2}$. The three output streams for $\mathbf{U}$, $\mathbf{\Sigma}$, and $\mathbf{V}$ are written in BRAMs C, D, and E, respectively. Since $\Sigma\in\mathbb{C}^{L-1\times K}$ is a diagonal matrix, with a large number of non-diagonal zeroes, the number of multiplications in equation \eqref{eq:svd_inv} can be significantly reduced. This architecture is designed for $K=2$. The simplified product $\mathbf{V}\mathbf{\Sigma}^+\mathbf{U}^H$ is highlighted in Figure \ref{fig:esprit_arch}(b) and is written in BRAM F. Despite these optimizations, this implementation has higher latency, since the SVD processing is on a large sized $(L-1)\times (L-1)$ square matrix. We refer to this implementation as the high-complexity ESPRIT.

\subsubsection{Novel low complexity pseudo inverse}
\begin{figure}[h]
    \centering
     \includegraphics[scale = 0.575]{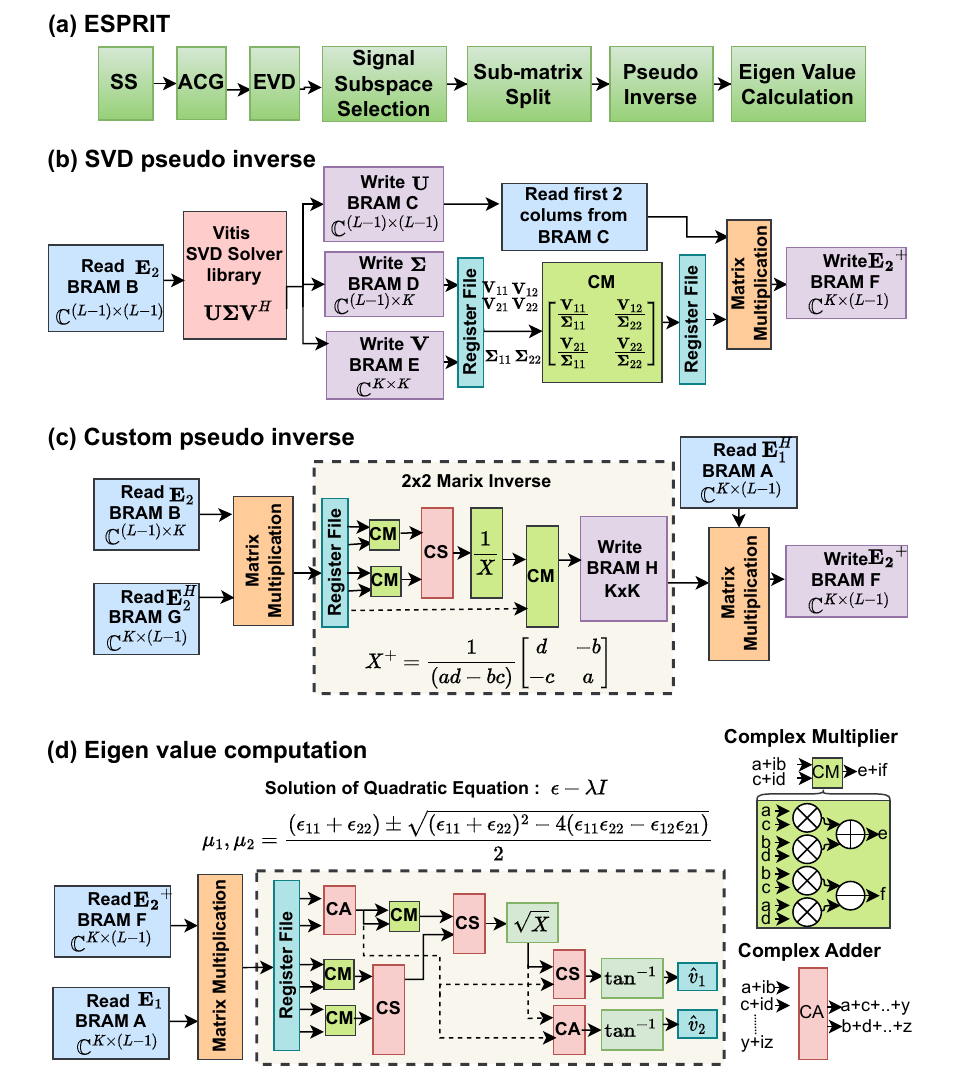}
     \vspace{-0.5cm}
    \caption{(a) Hardware blocks in ESPRIT, detailed hardware architecture of pseudo inverse with (b) SVD, (c) Proposed low-complexity implementation, and (d) eigen value computation.}
    \label{fig:esprit_arch}
\end{figure}

This is discussed in equation \eqref{eq:custom_inv} of Section \ref{Sec:Doppler_algo}, and presented in Figure \ref{fig:esprit_arch}(c). A copy of matrix $\mathbf{E_2}$ is stored in BRAM G, to explore BRAM partitioning in $\mathbf{E_2}^H\mathbf{E_2}$ matrix multiplication. The product is a $K\times K$ matrix, which is stored in the register file. This is followed by a $K\times K$ matrix inversion involving determinant and adjoint computation as shown in Figure \ref{fig:esprit_arch}(c). Unlike the SVD-based implementation, inversion on a square matrix with $K=2$ largely simplifies the pseudo inverse operation. The inverse matrix $(\mathbf{E_2}^H\mathbf{E_2})^+$ is written in BRAM H. This is followed by another matrix multiplication as shown in the Figure, and the final output is written in BRAM F. We refer to this implementation as the low-complexity ESPRIT. 

After the pseudo inverse computation, another matrix multiplication for the calculation of $\bm{\epsilon}\in \mathbb{C}^{K \times K}$ takes place as shown in equation \eqref{eq:epsilon_esprit}. This is discussed in Figure \ref{fig:esprit_arch}(d). $\bm{\epsilon}$ is written in registers, and this is followed by the eigen value calculation. For $K=2$, the implementation is simplified by finding the quadratic solution to $\bm{\epsilon}-\mu_kI$. The optimized eigenvalue computation architecture is presented in Figure \ref{fig:esprit_arch}(d), and the Doppler estimates are stored in registers.
\vspace{-0.2cm}
\subsection{Reconfigurable Doppler Estimation Architecture}
\begin{figure}[h]
    \centering
     \includegraphics[scale = 0.42]{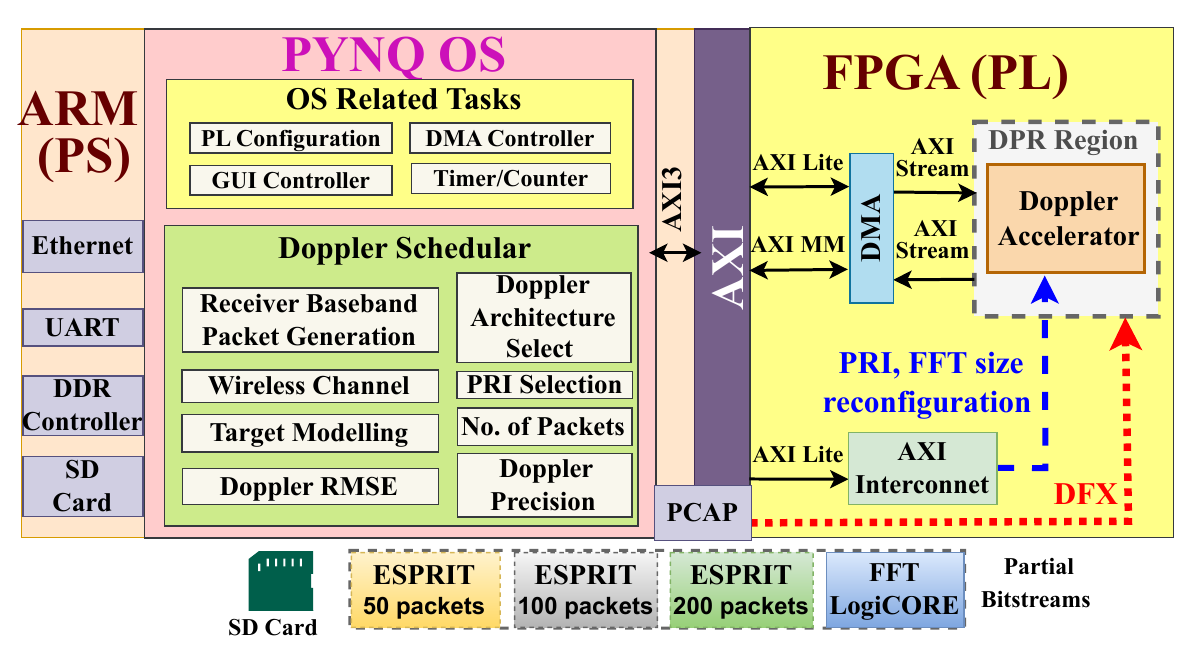}
     \vspace{-0.4cm}
\caption{Reconfigurable architecture for Doppler estimation via hardware software co-design on Zynq MPSoC}
    \label{fig:mpsoc_dpr}
\end{figure}

Zynq MPSoC comprises the quad-core A53 ARM processor referred to as the processing system (PS) and an Ultrascale FPGA referred to as the programmable logic (PL). We implement the Doppler estimation on the Zynq MPSoC using hardware-software co-design to partition tasks between the PS and PL, as shown in Figure \ref{fig:mpsoc_dpr}. Slow-time input generation, channel and target modeling, and performance analysis are conducted in PS, whereas Doppler processing is fully offloaded to PL. Data transfer between PS and PL occurs via direct memory access (DMA). The Doppler processing architecture is reconfigurable to switch between FFT and ESPRIT algorithms based on coarse or fine search requirements and can dynamically tune between different PRI, number of packets (CPI), and processing precisions. Reconfiguration control is provided by the processor via the AXI-Lite interface. Switching between algorithms is performed through dynamic partial reconfiguration (DPR) or dynamic function exchange (DFX). DFX is enabled through the processor configuration access port (PCAP), where the DPR region on the FPGA is reprogrammed by the chosen partial bitstream during run-time. The partial bitstreams for FFT and ESPRIT packets are stored on the SD card. The FFT implementation is realized using the built-in AMD Xilinx LogiCORE FFT IP, which supports reconfigurability between different FFT sizes from the PS during run-time.
\label{Sec:reconfig_arch}
\vspace{-0.4cm}

\section{Hardware Complexity and Performance Analysis}
\label{Sec:performance_analysis}
We evaluate the Doppler velocity estimation performance of the hardware IP cores in terms of root mean squared error (RMSE) under various signal-to-noise ratios (SNRs) in Rician channel conditions with Rician factor of 2 dB. We compare the performance and hardware complexity of FFT, MUSIC, and ESPRIT on the Zynq MPSoC. The FPGA mapping of MUSIC has been commonly explored in the literature \cite{tewari2024reconfigrsp, MUSIC_TCAS1, mansi2022spatialsensing}, and its hardware complexity and Doppler performance results are based on the implementation presented in \cite{tewari2024reconfigrsp}. We consider two scenarios for analysis on Doppler resolution.

\vspace{-0.2cm}
\subsection{Coarse Estimation}
\begin{figure}[!h]
    \centering
\includegraphics[scale = 0.35]{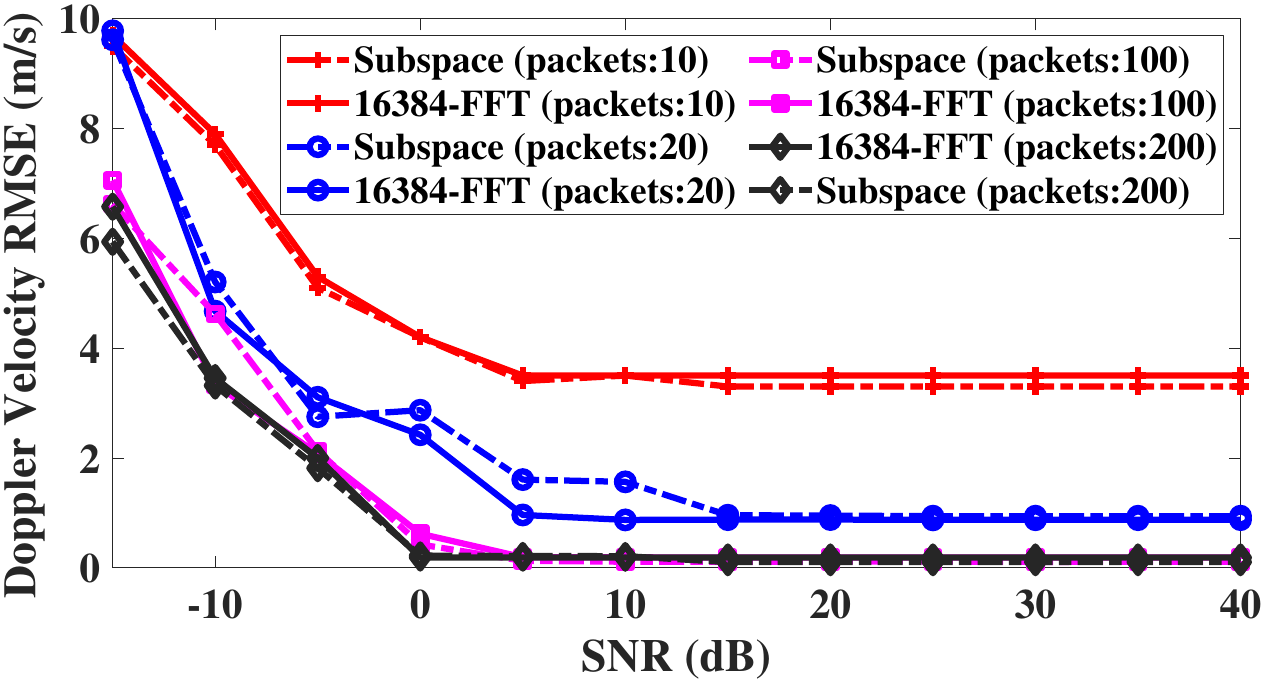}\vspace{-0.3cm}
    \caption{Doppler velocity RMSE comparison between FFT and subspace algorithms for different numbers of packets under single target detection}
    \label{fig:st_nop}
\end{figure}

\begin{figure}[!h]
    \centering
     \includegraphics[scale = 0.34]{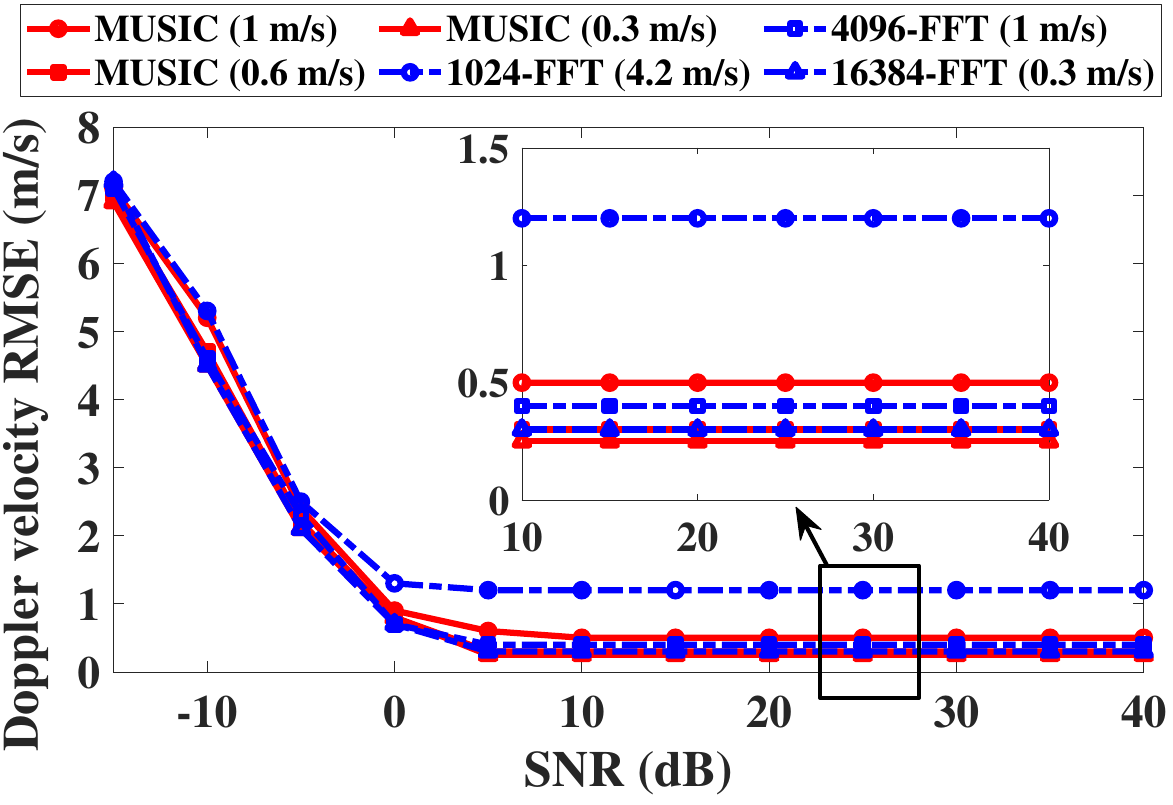}
     \vspace{-0.4cm}
    \caption{Doppler velocity RMSE with FFT and MUSIC for different Doppler precision with 100 packets}
    \label{fig:st_precision}
\end{figure}

If the presence of a target is detected in a range-azimuth bin, we perform the coarse estimation to identify it as MU or static clutter. In Figure~\ref {fig:st_nop}, we compare the RMSE for different numbers of packets over a wide range of SNRs. Here, we skipped the MUSIC since the performance of ESPRIT and MUSIC is identical for coarse estimation. 
As expected, RMSE improves with the increase in the number of packets. Further, the performance of FFT is nearly same as that of MUSIC/ESPRIT. In Figure~\ref{fig:st_precision}, we compare the RMSE performance for different Doppler precisions-$4.2m/s$, $1m/s$, $0.6m/s$, and $0.3m/s$, with 100 slow time packets. Even in this case, the performances of FFT and MUSIC are nearly identical. Here, ESPRIT is excluded because precision variation by changing the candidate Doppler search bins is only possible in MUSIC and FFT. As shown in  Table~\ref{tab:hwc_FFT_sizes}, FFT offers substantially lower complexity, power consumption, and latency than MUSIC. \emph{Thus, for coarse estimation, FFT is preferred due to its low complexity and 24 times faster execution time than the subspace algorithm.} Further, FFT size can be reconfigurable on-the-fly to further reduce execution time by lowering the precision, as shown in  Table~\ref{tab:hwc_FFT_sizes}.
\vspace{-0.3cm}

\vspace{-0.4cm}

\vspace{-0.6cm}
\begin{table}[!h]
\caption{Hardware Complexity for different Doppler precisions for FFT and subspace method on Zynq MPSoC with 100 packets. }
\label{tab:hwc_FFT_sizes}
\resizebox{0.49\textwidth}{!}{
\begin{tabular}{|c|c|cccc|c|c|}
\hline
\multirow{2}{*}{\textbf{Algorithm}} &
  \multirow{2}{*}{\textbf{\begin{tabular}[c]{@{}c@{}}Precision \\ (m/s)\end{tabular}}} &
  \multicolumn{4}{c|}{\textbf{Resource Utilization}} &
  \multirow{2}{*}{\textbf{\begin{tabular}[c]{@{}c@{}}Dynamic\\       Power (W)\end{tabular}}} &
  \multirow{2}{*}{\textbf{\begin{tabular}[c]{@{}c@{}}Latency\\      (ms)\end{tabular}}} \\ \cline{3-6}
                   &              & \multicolumn{1}{c|}{\textbf{BRAM}} & \multicolumn{1}{c|}{\textbf{LUT}} & \multicolumn{1}{c|}{\textbf{FF}} & \textbf{DSP} &      &      \\ \hline
\textbf{16384-FFT} & \textbf{0.3} & \multicolumn{1}{c|}{72.5}          & \multicolumn{1}{c|}{10872}        & \multicolumn{1}{c|}{15778}       & 18           & 3.5  & 1.18 \\ \hline
\textbf{4096-FFT}  & \textbf{1} & \multicolumn{1}{c|}{27.5}          & \multicolumn{1}{c|}{10559}        & \multicolumn{1}{c|}{15652}       & 18           & 3.46 & 0.56 \\ \hline
\textbf{1024-FFT}  & \textbf{4.2}   & \multicolumn{1}{c|}{18}            & \multicolumn{1}{c|}{10431}        & \multicolumn{1}{c|}{15544}       & 18           & 3.44 & 0.52 \\ \hline
\multirow{2}{*}{\textbf{MUSIC}} &
  \textbf{0.3} &
  \multicolumn{1}{c|}{\multirow{2}{*}{293.5}} &
  \multicolumn{1}{c|}{\multirow{2}{*}{60497}} &
  \multicolumn{1}{c|}{\multirow{2}{*}{60001}} &
  \multirow{2}{*}{483} &
  4.81 &
  28.6 \\ \cline{2-2} \cline{7-8} 
                   & \textbf{1}   & \multicolumn{1}{c|}{}              & \multicolumn{1}{c|}{}             & \multicolumn{1}{c|}{}            &              & 4.83 & 26.9 \\ \hline
\end{tabular}
}
\end{table}

\subsection{Fine Estimation with Tightly-Spaced MUs}
\begin{figure}[h]
    \centering
     \includegraphics[scale = 0.32]{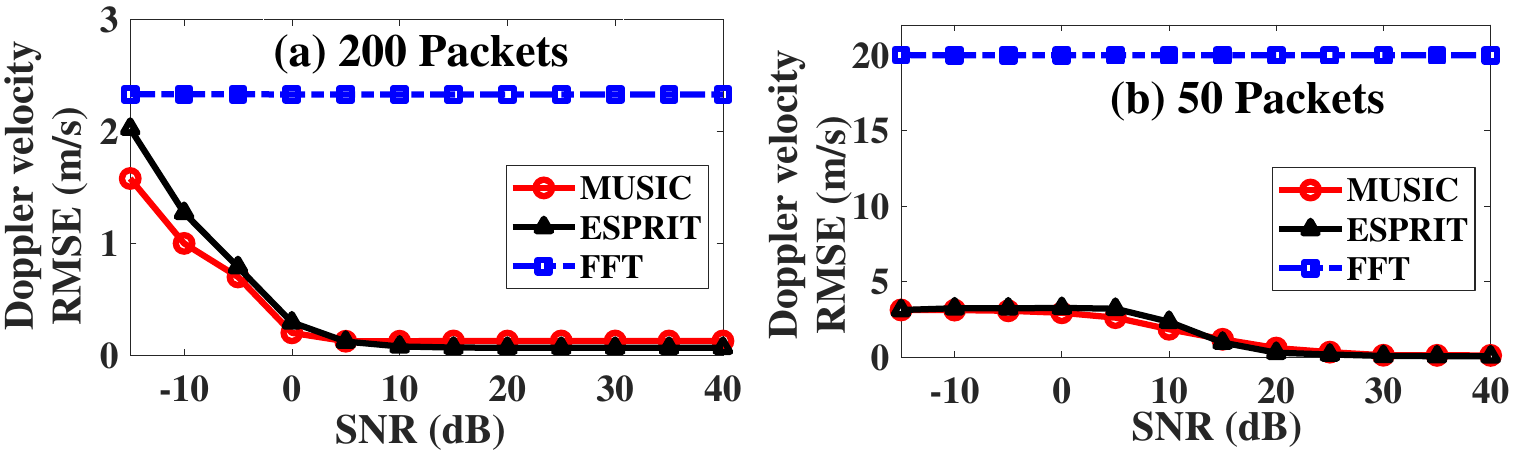}
    \caption{Doppler RMSE comparison of FFT, MUSIC, and low-complexity ESPRIT for two targets separated by 6 m/s using a 2 µs PRI with (a) 200 and (b) 50 packets}
    \label{fig:mt_cmp}
\end{figure}

For every range-azimuth bin, if the presence of a target is detected, fine estimation is used for three tasks: 1) Distinguish between MUs and static clutter, 2) Estimate the number of MUs, and 3) Estimate the Doppler velocity of each MU. Here, we assume that the number of MUs with identical range and azimuth is at most 2. In Figure.~\ref{fig:mt_cmp}, we compare the average Doppler velocity RMSE for FFT, ESPRIT, and MUSIC algorithms. Here, we assume 200 packets with a velocity difference of 6$m/s$ and a PRI of 2 $\mu s$. It can be observed that the FFT fails to distinguish between multiple MUs, resulting in higher RMSE. Among MUSIC and ESPRIT, the proposed ESPRIT offers slightly better performance at lower SNRs. The performance of the low and high-complexity ESPRIT is identical; hence, a single ESPRIT plot is indicated in black colour. To improve the FFT performance, the number of packets must be higher than 2048, which is not practically feasible for ISAC, as the extended radar cycle will effectively reduce the communication cycle duration to zero.  

\begin{table*}[!t]
\caption{Hardware Complexity comparison for MUSIC, low-complexity and high-complexity ESPRIT for 50 and 200 packets}
\Huge
\label{tab:hwc_music_esprit}
\vspace{-0.2cm}
\centering
\renewcommand{\arraystretch}{1}
\resizebox{1\textwidth}{!}{
\begin{tabular}{|c|ccccccc|ccccccc|}
\hline
\textbf{Packets} &
  \multicolumn{7}{c|}{\textbf{50}} &
  \multicolumn{7}{c|}{\textbf{200}} \\ \hline
\multirow{2}{*}{\textbf{\begin{tabular}[c]{@{}c@{}}Subspace \\  Algorithm\end{tabular}}} &
  \multicolumn{4}{c|}{\textbf{Resource Utilization}} &
  \multicolumn{1}{c|}{\multirow{2}{*}{\textbf{\begin{tabular}[c]{@{}c@{}}Dynamic\\       Power (W)\end{tabular}}}} &
  \multicolumn{1}{c|}{\multirow{2}{*}{\textbf{\begin{tabular}[c]{@{}c@{}}Latency  \\ (ms)\end{tabular}}}} &
  \multirow{2}{*}{\textbf{\begin{tabular}[c]{@{}c@{}}AF  \\ w.r.t PS\end{tabular}}} &
  \multicolumn{4}{c|}{\textbf{Resource Utilization}} &
  \multicolumn{1}{c|}{\multirow{2}{*}{\textbf{\begin{tabular}[c]{@{}c@{}}Dynamic\\       Power (W)\end{tabular}}}} &
  \multicolumn{1}{c|}{\multirow{2}{*}{\textbf{\begin{tabular}[c]{@{}c@{}}Latency   \\ (ms)\end{tabular}}}} &
  \multirow{2}{*}{\textbf{\begin{tabular}[c]{@{}c@{}}AF \\ w.r.t PS\end{tabular}}} \\ \cline{2-5} \cline{9-12}
 &
  \multicolumn{1}{c|}{\textbf{BRAM}} &
  \multicolumn{1}{c|}{\textbf{LUT}} &
  \multicolumn{1}{c|}{\textbf{FF}} &
  \multicolumn{1}{c|}{\textbf{DSP}} &
  \multicolumn{1}{c|}{} &
  \multicolumn{1}{c|}{} &
   &
  \multicolumn{1}{c|}{\textbf{BRAM}} &
  \multicolumn{1}{c|}{\textbf{LUT}} &
  \multicolumn{1}{c|}{\textbf{FF}} &
  \multicolumn{1}{c|}{\textbf{DSP}} &
  \multicolumn{1}{c|}{} &
  \multicolumn{1}{c|}{} &
   \\ \hline
\textbf{\begin{tabular}[c]{@{}c@{}}ESPRIT \\      High complexity\end{tabular}} &
  \multicolumn{1}{c|}{55} &
  \multicolumn{1}{c|}{88551} &
  \multicolumn{1}{c|}{102296} &
  \multicolumn{1}{c|}{335} &
  \multicolumn{1}{c|}{4.99} &
  \multicolumn{1}{c|}{73.39} &
  1.46 &
  \multicolumn{1}{c|}{533} &
  \multicolumn{1}{c|}{110675} &
  \multicolumn{1}{c|}{113403} &
  \multicolumn{1}{c|}{386} &
  \multicolumn{1}{c|}{5} &
  \multicolumn{1}{c|}{1887.4} &
  1.22 \\ \hline
\textbf{\begin{tabular}[c]{@{}c@{}}ESPRIT\\      Low complexity\end{tabular}} &
  \multicolumn{1}{c|}{\begin{tabular}[c]{@{}c@{}}11   \\ (-80\%)\end{tabular}} &
  \multicolumn{1}{c|}{\begin{tabular}[c]{@{}c@{}}44144   \\ (-50.2\%)\end{tabular}} &
  \multicolumn{1}{c|}{\begin{tabular}[c]{@{}c@{}}57053   \\ (-44.2\%)\end{tabular}} &
  \multicolumn{1}{c|}{\begin{tabular}[c]{@{}c@{}}232   \\ (-30.8\%)\end{tabular}} &
  \multicolumn{1}{c|}{\begin{tabular}[c]{@{}c@{}}4.30   \\ (-13.8\%)\end{tabular}} &
  \multicolumn{1}{c|}{\begin{tabular}[c]{@{}c@{}}15.73   \\ (-78.6\%)\end{tabular}} &
  1.86 &
  \multicolumn{1}{c|}{\begin{tabular}[c]{@{}c@{}}81   \\ (-84.8\%)\end{tabular}} &
  \multicolumn{1}{c|}{\begin{tabular}[c]{@{}c@{}}49926   \\ (-54.9\%)\end{tabular}} &
  \multicolumn{1}{c|}{\begin{tabular}[c]{@{}c@{}}61236   \\ (-46\%)\end{tabular}} &
  \multicolumn{1}{c|}{\begin{tabular}[c]{@{}c@{}}267   \\ (-30.8\%)\end{tabular}} &
  \multicolumn{1}{c|}{\begin{tabular}[c]{@{}c@{}}4.3   \\ (-14\%)\end{tabular}} &
  \multicolumn{1}{c|}{\begin{tabular}[c]{@{}c@{}}30.1   \\ (-98.4\%)\end{tabular}} &
  28.25 \\ \hline
\textbf{MUSIC} &
  \multicolumn{1}{c|}{\begin{tabular}[c]{@{}c@{}}243.5   \\ (+340\%)\end{tabular}} &
  \multicolumn{1}{c|}{\begin{tabular}[c]{@{}c@{}}44305   \\ (-50\%)\end{tabular}} &
  \multicolumn{1}{c|}{\begin{tabular}[c]{@{}c@{}}40980   \\ (-60\%)\end{tabular}} &
  \multicolumn{1}{c|}{\begin{tabular}[c]{@{}c@{}}359   \\ (+7.2\%)\end{tabular}} &
  \multicolumn{1}{c|}{\begin{tabular}[c]{@{}c@{}}4.6   \\ (-7.8\%)\end{tabular}} &
  \multicolumn{1}{c|}{\begin{tabular}[c]{@{}c@{}}7.6   \\ (-89.6\%)\end{tabular}} &
  193.54 &
  \multicolumn{1}{c|}{\begin{tabular}[c]{@{}c@{}}394.5   \\ (-26\%)\end{tabular}} &
  \multicolumn{1}{c|}{\begin{tabular}[c]{@{}c@{}}96577   \\ (-12.7\%)\end{tabular}} &
  \multicolumn{1}{c|}{\begin{tabular}[c]{@{}c@{}}94252   \\ (-17\%)\end{tabular}} &
  \multicolumn{1}{c|}{\begin{tabular}[c]{@{}c@{}}723   \\ (+46.6\%)\end{tabular}} &
  \multicolumn{1}{c|}{\begin{tabular}[c]{@{}c@{}}5.8   \\ (+16\%)\end{tabular}} &
  \multicolumn{1}{c|}{\begin{tabular}[c]{@{}c@{}}201.13   \\ (-89.3\%)\end{tabular}} &
  11.1 \\ \hline
\end{tabular}
}
\end{table*}
\vspace{-0.1cm}
Table~\ref{tab:hwc_music_esprit} compares the complexity, power consumption, and execution time of ESPRIT and MUSIC architectures on Zynq MPSoC for 50 and 200 packets. We consider MUSIC architecture in \cite{tewari2024reconfigrsp}, high and low complexity ESPRIT, as discussed in Section~\ref{Sec:Doppler_algo}. The proposed low-complexity ESPRIT implementation provides up to 80\% and 30\% savings in BRAM and DSP, respectively, 14\% reduction in dynamic power, and 98\% reduction in latency over the high complexity implementation without any compromise on functional accuracy. \emph{When compared to MUSIC, the low-complexity ESPRIT provides significant savings in area (BRAM:-79.5\%, LUT:-48.3\%, FF:-35\%, DSP:-63\%), up to 26\% reduction in power, and 6.7 times faster execution time, making it the preferred super-resolution algorithm for fine search.} We also compare the execution time of these algorithms in terms of acceleration factor (AF), which is the ratio of the execution times on ARM and FPGA. We observe that the proposed low complexity ESPRIT offers significantly higher AF of 28.25, indicating its suitability for parallel processing on FPGA or ASIC. 

\begin{figure}[!h]
    \centering
     \includegraphics[scale = 0.35]{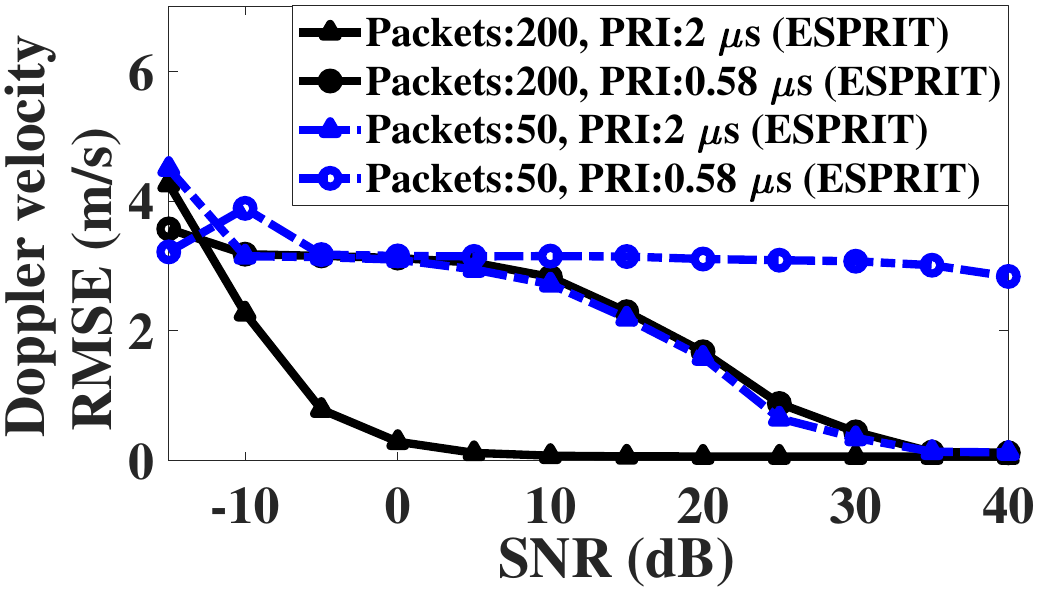}
     \vspace{-0.4cm}
    \caption{PRI (0.58$\mu s$ and 2$\mu s$) selection for multiple targets: resolving two targets with 6 m/s velocity difference with 50 and 200 packets using ESPRIT. 
    }
    \label{fig:mt_PRI}
\end{figure}

\begin{figure*}[t]
\centering
\includegraphics[scale = 0.38]{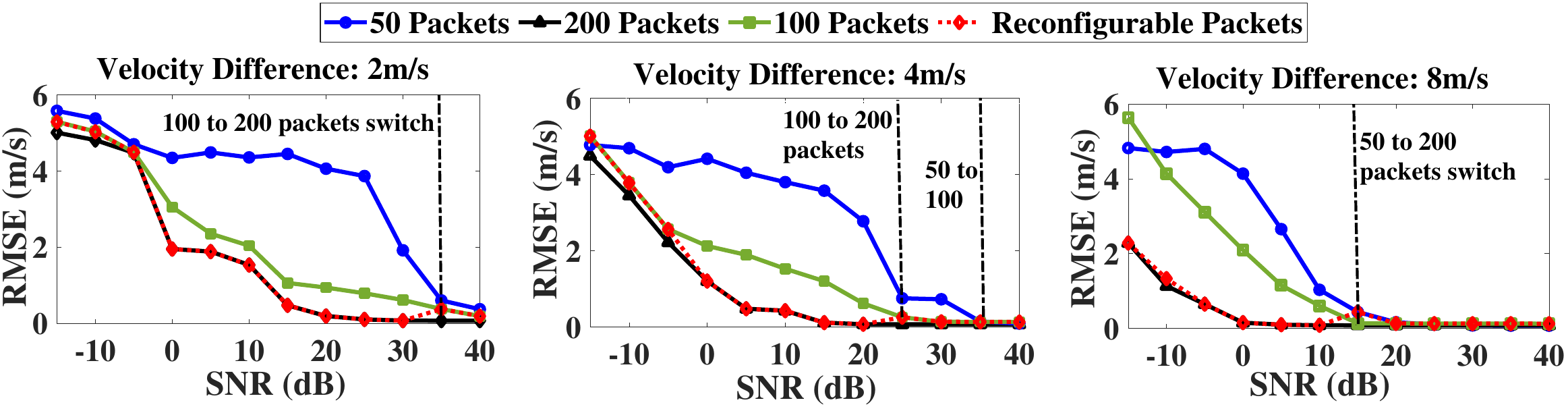}
     \vspace{-0.4cm}
    \caption{Doppler velocity RMSE with ESPRIT for the detection of two targets with varying velocity difference }
\label{fig:mt_veldiff}
\end{figure*}

The Doppler resolution can be improved by two approaches: increasing the number of packets or the PRI. Here, the latter approach also impacts the radar's maximum unambiguous range and Doppler velocity. Figure~\ref {fig:mt_PRI} shows that for a fixed number of packets, the Doppler RMSE can be significantly reduced by increasing the PRI from $0.58\mu s$ to $2\mu s$. It shows that for a CPI of $100 \mu s$, both 200 packets with $0.58\mu s$ PRI and 50 packets with $2\mu s$ PRI offer identical performance in resolving two targets spaced 6 $m/s$ apart. Here, the latter (2$\mu s$) PRI is a preferable choice due to the low hardware complexity of ESPRIT with 50 packets, as shown in Table~\ref{tab:hwc_music_esprit}. For resolving even finer velocity differences between two targets, the approach of increasing PRI beyond $2 \mu s$ is no longer feasible due to a very large number of fast-time samples required in matched filtering, resulting in on-chip memory (BRAM) overshoot on the FPGA. Additionally, this also reduces the maximum unambiguous Doppler, which affects the detection of fast-moving targets. Thus, now the preferred approach is to increase the packets for finer resolution. Figure \ref{fig:mt_veldiff} shows that targets with even finer velocity differences, up to 2 $m/s$, can be resolved by increasing the number of packets to 200 with PRI $2 \mu s$. Table \ref{tab:hwc_esprit} compares the hardware complexity for different numbers of packets with PRI 2$\mu s$, showing increased complexity and improved performance with higher packets/CPI. From Figure \ref{fig:mt_veldiff}, it can be inferred that switching to the 50 packet ESPRIT under high SNR conditions can provide up to 2$\times$ improvement in latency over 200 packets (shown in Table \ref{tab:hwc_esprit}), without any drop in performance. In Figure \ref{fig:mt_veldiff}, the performance of the reconfigurable Doppler architecture is indicated with a red dotted line, and the SNR-based switching points from low to high numbers of packets are indicated with a black vertical line for velocity differences of 2, 4, and 8 m/s. \emph{Thus, for resolving multiple closely spaced targets, the reconfigurable Doppler architecture can on-the-fly tune to different packet numbers, based on channel conditions and accuracy requirements, maintaining optimum performance.}  
 
 
\section{Conclusion}
\vspace{-0.2cm}
\label{Sec:conclusion}
This work presents a run-time reconfigurable Doppler processing architecture that switches between different algorithms (FFT and ESPRIT) and RSP parameters (number of input packets, PRI, and precision) based on ISAC search accuracy requirements. From the performance and hardware complexity analysis, FFT is preferred for coarse search, whereas ESPRIT is preferred for fine search with multiple target localization. The low-complexity ESPRIT provides 79\% and 63\% BRAM and DSP savings, and a 6.7$\times$ faster FPGA execution time compared to MUSIC. Further, dynamic packet selection in ESPRIT improves processing latency by 2$\times$ under high-SNR conditions. In the future, we will explore hardware architectures for deep-learning–augmented Doppler estimation to enhance Doppler performance.

\begin{table}[!t]
\caption{Hardware Complexity comparison ESPRIT for different number of packets with PRI 2 $\mu s$}
\vspace{-0.2cm}
\label{tab:hwc_esprit}
\centering
\renewcommand{\arraystretch}{1.1}
\resizebox{0.42
\textwidth}{!}{
\begin{tabular}{|c|c|cccc|c|}
\hline
\multirow{2}{*}{\textbf{\begin{tabular}[c]{@{}c@{}}CPI \\ \textbf{($\mu s$)}\end{tabular}}} &
  \multirow{2}{*}{\textbf{Packets}} &
  \multicolumn{4}{c|}{\textbf{Resource Utilization}} &
  \multirow{2}{*}{\textbf{\begin{tabular}[c]{@{}c@{}}Latency\\  \textbf{(ms)}\end{tabular}}} \\ \cline{3-6}
    &     & \multicolumn{1}{c|}{\textbf{BRAM}} & \multicolumn{1}{c|}{\textbf{LUT}} & \multicolumn{1}{c|}{\textbf{FF}} & \textbf{DSP} &       \\ \hline
100 &
  50 &
  \multicolumn{1}{c|}{\begin{tabular}[c]{@{}c@{}}11\\ (-86.4\%)\end{tabular}} &
  \multicolumn{1}{c|}{\begin{tabular}[c]{@{}c@{}}44144 \\ (-11.5\%)\end{tabular}} &
  \multicolumn{1}{c|}{\begin{tabular}[c]{@{}c@{}}57053 \\ (-6.8\%)\end{tabular}} &
  \begin{tabular}[c]{@{}c@{}}232 \\ (-13\%)\end{tabular} &
  \begin{tabular}[c]{@{}c@{}}15.72  \\ (-47.8\%)\end{tabular} \\ \hline
200 &
  100 &
  \multicolumn{1}{c|}{\begin{tabular}[c]{@{}c@{}}31 \\ (-61.7\%)\end{tabular}} &
  \multicolumn{1}{c|}{\begin{tabular}[c]{@{}c@{}}44070 \\ (-11.7\%)\end{tabular}} &
  \multicolumn{1}{c|}{\begin{tabular}[c]{@{}c@{}}57219\\ (-6.6\%)\end{tabular}} &
  \begin{tabular}[c]{@{}c@{}}240\\ (-10\%)\end{tabular} &
  \begin{tabular}[c]{@{}c@{}}19.4\\ (-35.5\%)\end{tabular} \\ \hline
400 & 200 & \multicolumn{1}{c|}{81}            & \multicolumn{1}{c|}{49926}        & \multicolumn{1}{c|}{61236}       & 267          & 30.1 \\ \hline
\end{tabular}
}
\end{table}
\bibliographystyle{IEEEtran}
\bibliography{References}
\end{document}